# Tuning stoichiometry and its impact on superconductivity of monolayer and multilayer FeSe on SrTiO$_3$

Chong Liu,[1,2] and Ke Zou[1,2,*]

[1] *Quantum Matter Institute, University of British Columbia, Vancouver, British Columbia V6T 1Z4, Canada*

[2] *Department of Physics and Astronomy, University of British Columbia, Vancouver, British Columbia V6T 1Z1, Canada*

[*]kzou@phas.ubc.ca

(Dated: February 14, 2020)

Synthesis of monolayer FeSe on SrTiO$_3$, with greatly enhanced superconductivity compared to bulk FeSe, remains difficult. Lengthy annealing within a certain temperature window is always required to achieve superconducting samples as reported by different groups around the world, but the mechanism of annealing in inducing superconductivity has not been elucidated. We grow FeSe films on SrTiO$_3$ by molecular beam epitaxy and adjust the stoichiometry by depositing additional small amounts of Fe atoms. The monolayer films become superconducting after the Fe deposition without any annealing, and show similar superconducting transition temperatures as those of the annealed films in transport measurements. We also demonstrate on the 5-unit-cell films that the FeSe multilayer can be reversibly tuned between the non-superconducting $\sqrt{5} \times \sqrt{5}$ phase with Fe-vacancies and superconducting $1 \times 1$ phase. Our results reveal that the traditional anneal process in essence removes Fe vacancies and the additional Fe deposition serves as a more efficient way to achieve superconductivity. This work highlights the significance of stoichiometry in the superconductivity of FeSe thin films and provides an easy path for superconducting samples.

FeSe, with the simplest structure and chemical composition among Fe-based superconductors, has been intensely studied in the past decade [1]. It distinguishes itself in various aspects and poses many remaining unresolved questions. Bulk tetragonal FeSe has a superconducting transition temperature $T_c \approx 8$ K [2]. Monolayer tetragonal FeSe grown on SrTiO$_3$ (STO) show greatly interface-enhanced superconductivity with $T_c$ up to 65 K [3-6], albeit only after a long post-annealing process. Efforts have been focused on understanding the mechanisms of enhanced superconductivity, since this may lead the way to the design of new superconductors with higher $T_c$.

High $T_c$ superconducting FeSe thin films are mostly grown by molecular beam epitaxy (MBE). A Se-rich condition is employed with the flux ratio $\Phi_{Se}$:$\Phi_{Fe}$ ranging from 5 to 20, with a substrate temperature well above the evaporation temperature of Se [6-8]. Ideally, FeSe grown this way should be stoichiometric with Fe:Se = 1:1, but the long post-growth annealing process required to reach superconducting states suggests otherwise [9,10]. Although stoichiometry has been difficult to characterize in thin film FeSe, it does affect the properties of bulk FeSe substantially. There exist multiple bulk phases of FeSe such as hexagonal $\alpha$-Fe$_x$Se, tetragonal $\beta$-Fe$_x$Se and hexagonal $\gamma$-Fe$_7$Se$_8$ [11,12], among which only the $\beta$ phase is superconducting when $x \approx 1$ [12,13]. Almost defect-free $\beta$-FeSe single crystals have been obtained with $T_c \sim 10$ K [14,15]. Depending on the synthesis conditions, Fe vacancies or excess Fe atoms could be present in the bulk crystals and suppress the superconductivity. Substantially more Fe vacancies in $\beta$-Fe$_x$Se leads to ordered crystal forms such as Fe$_3$Se$_4$, Fe$_4$Se$_5$, Fe$_9$Se$_{10}$ [16].

Because of the high Se flux used in MBE growth, the existence of excess Fe atoms or Se vacancies in thin FeSe films become unlikely. The presence of Fe vacancies may lead to the as-grown non-superconducting thin films. By *in situ* scanning tunneling microscopy (STM), the as-grown FeSe films on STO show high density of dumbbell-like defects and an insulating energy gap. With annealing, the defects are reduced, the film becomes metallic and the superconducting gap opens up at Fermi energy [7,9,17,18]. In fact, similar observation has been made on FeSe films on graphene as well, although the superconducting gaps are different [8,18]. Density functional theory calculations have attributed these dumbbell defects to Fe vacancies and claimed that they can diffuse between Fe sites during annealing and eventually move to the edge of the film [18]. Opposite observations include a recent STM work revealing increase of FeSe film coverage after depositing additional Se at a low temperature, suggesting ~20% excess Fe in their films [19]. In addition, one scanning transmission electron microscope work shows the existence of interstitial Se atoms between FeSe and STO, although not in the FeSe itself [20], raising further questions about stoichiometry in few-layer FeSe films. Lack of direct experimental evidence makes it difficult to pin down how the stoichiometry affects the superconducting transition in FeSe films.

In this work, we grow few unit-cell (UC) FeSe films on STO and deposit small amounts of additional Fe onto the as-grown films. Adding the Fe atoms substantially improves the film quality and induces superconductivity without post-anneal process. This approach serves as an alternative, easy, and time-saving method compared to the many hours of annealing required in the past. The $T_c$ of monolayer FeSe on STO by this method can reach as high as the annealed samples in transport measurements, directly indicating that the as-grown FeSe films possess Fe vacancies in the layer structure. In a 5-UC sample, we can control the surface reconstruction and stoichiometry by alternating deposition of Se and Fe and switch the FeSe multilayer between a Fe-vacancy phase and the superconducting phase.

FeSe films were grown on treated undoped STO substrates (CrysTec GmbH), using procedures similar to other groups [9,21,22]. The base pressure of the MBE chamber was ~1 × 10$^{-10}$ Torr. The flux ratio $\Phi_{Se}:\Phi_{Fe} \approx 5$. The substrate temperature was 420 °C for growth and 480 °C for anneal. The samples were monitored by reflection high-energy electron diffraction (RHEED). The growth rate was ~0.2 UC/minute, determined by RHEED intensity oscillation [23]. RHEED pattern simulations were generated by a program written in Wolfram Mathematica. For transport measurements, samples were capped by 12-UC FeTe grown at 280 °C following standard procedures [21]. Transport measurements were carried out in a Quantum Design physical property measurement system (PPMS) with four-terminal method. Gold wires were bonded on the samples with indium lumps as the electrodes.

Figures 1(a) and (b) show the typical RHEED images of STO and as-grown monolayer FeSe grown at 420 °C, respectively. The FeSe film has 1 × 1 square structure with lattice constant $a$ = 3.85 angstroms, calibrated by the distance between diffraction streaks. Unlike the usual routine of annealing at higher temperature in ultrahigh vacuum for many hours, we kept the film at the growth temperature 420 °C, and deposited Fe atoms onto it. The RHEED intensity increased as shown in Figs. 1(c) and (d). It reached the maximum in 37 seconds and then decreased and we stopped the Fe deposition after that (shaded area in Fig. 1(d)). Both (01) and (00) diffraction streak intensities showed similar trend with deposition especially the (01) was increased by 24 %. The RHEED intensity was enhanced without any sign of new phases or reconstructions, indicating that the as-grown film contained disordered Fe-vacancies and the subsequently deposited Fe improved the stoichiometry and the overall quality of the film. Once the optimal stoichiometry was reached, which corresponded to the maximum intensity, additional Fe atoms started to accumulate on the surface and lower the RHEED intensity.

The extreme sensitivity of RHEED to the surface crystal structures allows us to capture small changes in the amount of Fe vacancies in the film. Taking the maximum RHEED intensive peaks as the indication of correct stoichiometry, we estimate the Fe vacancy fraction $\delta$ in as-grown $Fe_{1-\delta}Se$ film by the deposition time $\delta = t_{Fe}/(t_{FeSe}+t_{Fe}) = 0.12$, where $t_{FeSe}$ and $t_{Fe}$ are the Fe-Se codeposition time and the Fe deposition time before the maximal intensity (Fig. 1(d)), respectively, with the same Fe flux. The estimated Fe vacancy fraction $\delta$ varies between different monolayer samples and ranges from 0.07 to 0.16 in our experiments.

The films were cooled down to 280 °C right after the Fe deposition and capped with 12-UC FeTe. Fig. 1(e) shows the transport measurements results of a typical as-grown sample and the Fe-adjusted sample. The as-grown sample only showed the typical structural and antiferromagnetic transition of the FeTe capping layer at 80 K [24] and an insulating behaviour below 15 K, while the Fe-adjusted sample was more conductive and exhibited a superconducting transition. Closer observation in low-temperature range shown in Fig. 1(f) gives the transition temperatures $T_c^{onset}$ = 31 K, $T_c^{mid}$ = 23 K, and $T_c^0$ = 18 K, similar to other groups' results of annealed samples [9,10,20]. The $T_c$ shifted to lower temperatures when an external magnetic field was applied, a result of the Meissner effect in a superconductor. We conclude that enhanced superconductivity in monolayer FeSe/STO can be achieved merely by increasing Fe content.

To compare, we prepared another monolayer sample (Figure 2) and applied the standard annealing process at 480 °C for 3 hours. As shown in Figs. 2(a) and (b), the RHEED intensity increased after annealing, manifesting the improvement of the crystal quality. To calculate the Fe vacancy content in annealed samples, we deposited Fe at 420 °C and carried out the same analysis method used in Fig. 1(d). The time to reach the maximal RHEED intensity was only 8 seconds and the relative increase of the intensity was < 5 % (red curve in Fig. 2(c)), much less than those for the as-grown samples (black curve in Fig. 2(c)), indicating the annealed film contained lower Fe vacancy fraction ($\delta \approx 0.02$). As shown in Fig. 2(d), the transport results are quite similar to those of the merely Fe-adjusted samples, with $T_c^{onset}$ = 33 K, $T_c^{mid}$ = 23 K and $T_c^0$ = 16 K.

We have shown both annealing and Fe deposition can induce superconductivity in FeSe films with similar transition temperatures. Either method reduces the Fe vacancies that are commonly generated during the MBE growth procedures, and eventually leads to stoichiometric films. However, adding Fe atoms is more efficient compared to the typical several to tens of hours annealing that is required.

This approach is further confirmed by RHEED studies of multilayer films. Reconstructed surface structures can provide a good indication of the existence of defects such as Fe vacancies, and their evolution can be easily tracked and characterized by RHEED. We utilize the ($\sqrt{5} \times \sqrt{5}$)-$R26.6°$ reconstruction that can be generated on thicker FeSe films to guide our experiments on multilayer films. Heavily Se-rich multilayer FeSe may show either $\sqrt{5} \times \sqrt{5}$ or $\sqrt{5} \times \sqrt{10}$ reconstruction [7,8,25], and the corresponding films are both insulating. STM studies have shown that the $\sqrt{5} \times \sqrt{5}$ reconstructions consists of ordered dumbbell-like defects which are suggested to be Fe vacancies by theoretical calculations [8,18].

We deposited a 5-UC FeSe film that was $1 \times 1$ as grown, and then intentionally introduced ($\sqrt{5} \times \sqrt{5}$)-$R26.6°$ reconstruction by annealing the film under Se flux at 250 °C for 5 mins (Fig. 3 and Fig. 4(a)). Note that the 250 °C temperature is above the evaporation temperature of Se, so most of Se will re-evaporate from the surface, while a small number of Se atoms diffuse into the film and drive the whole film to the $\sqrt{5} \times \sqrt{5}$ structure with more Fe vacancies. As shown in Figs. 3(a) and (b), new diffraction streaks emerge in addition to the original $1 \times 1$ FeSe. By assuming a $1 \times 1$ square lattice with two $\sqrt{5} \times \sqrt{5}$ reconstructions which are rotated by ±26.6° with respect to the $1 \times 1$ lattice, the computer simulations of the RHEED patterns (Figs. 3(c) and (d)) agree well with the experimental results in both [100] and [110] directions. One candidate for the $\sqrt{5} \times \sqrt{5}$ structure is the Fe$_4$Se$_5$ phase, in which 20% Fe vacancies are arranged in $\sqrt{5} \times \sqrt{5}$ ordering [16]. As depicted in Fig. 3(e), the unit cell of the vacancy-ordered Fe$_4$Se$_5$ lattice is consistent with $(\sqrt{5} \times \sqrt{5})$-$R26.6°$ supercell of FeSe and naturally has two equivalent orientations that lead to two kinds of domains in the films. Interestingly, Fe$_4$Se$_5$ is an antiferromagnetic insulator [16], also consistent with recent observations on as-grown multilayer FeSe films on STO [26].

Depositing Fe onto this $\sqrt{5} \times \sqrt{5}$ surface can recover the structure to $1 \times 1$ FeSe (Figs. 4(a) and (b)). The (01) diffraction streak intensity increased by ~60% when reaching the maximum at $t_{Fe}$ = 364 seconds (Fig. 4(c)) and the $\sqrt{5} \times \sqrt{5}$ order disappeared during the deposition (Fig. 4(b)). Note that the deposition time $t_{Fe}$ here is much longer than that for the 1-UC films (Fig. 1(d)), indicating that the Fe vacancies existed not only in the top layer but in the whole film, and the Fe atoms diffused into the bottom layers. With $t_{FeSe}$ = 25 min for the initial growth of 5-UC Fe$_{1-\delta}$Se film, $\delta = t_{Fe}/(t_{FeSe}+t_{Fe}) = 0.195$, close to the vacancy density in Fe$_4$Se$_5$ ($\delta$ = 0.2).

Intriguingly, this process is reversible by alternating the deposition of Fe and Se atoms. We carried out cycles of Se and Fe deposition on the same sample (Fig. 4(c)) and found that the film switched between two phases: the $\sqrt{5} \times \sqrt{5}$ order as shown in Fig. 4(a) after Se deposition and $1 \times 1$ order as shown in Fig. 4(b) after Fe deposition. The 5-UC film after two cycles of Se-Fe deposition still showed superconductivity in transport measurement with $T_c^{mid}$ = 18 K (Fig. 4(d)). The possible causes of the lower $T_c$ include (1) the $T_c$ of FeSe on STO decreases with increasing film thickness [10]; (2) for multilayer FeSe on STO, only the first layer can be superconducting [3,27], and it's more difficult for Fe atoms to penetrate the top layers to completely restore the stoichiometry of the bottom layers. Nevertheless, the results of the 5-UC film show that our method can control the stoichiometry of FeSe in a straightforward fashion, and demonstrate that the stoichiometry of FeSe films plays a key role in the emergence of superconductivity.

Figure 5 summarizes transport data of multiple 1-UC and 2-UC FeSe films on STO. The as-grown film is not superconducting, while all the other films, whether treated by anneal or Fe deposition or a combination of both, share similar superconducting transition temperatures. Although the transition widths differ (most likely due to different homogeneity of the films), the $T_c^{mid}$, which reflects the average $T_c$ of the entire sample, falls in a narrow range of 22–24 K. This again manifests that the essence of the annealing process is increasing the Fe content in Fe$_x$Se toward $x$ = 1. We not that $x$ could be above 1 if Fe atoms are over-deposited. However, the excess Fe on FeSe films, if any, will react with Te during the growth of FeTe capping layers. As magnetic impurities, excess Fe atoms would strongly suppress the superconductivity [28]. Therefore, the superconducting FeSe films showing optimal $T_c$, either annealed or Fe-adjusted, should be close to stoichiometry.

It has been widely accepted that electrons transferred from the STO to the FeSe film play a critical role in determining the superconducting properties [4,29]. Oxygen vacancies dope the STO surface and provide electrons for FeSe films [30]. They are usually generated during the annealing of STO in vacuum [22,31]. Since we have made the FeSe films superconducting without long time annealing, we argue that the oxygen vacancies are not primarily generated in the annealing process but in the pre-treatments of the STO substrates. Hence, the increase of electron concentration observed in FeSe films during annealing [4] probably result from the removal of Fe vacancies.

Various mechanisms might contribute to the restoration of superconductivity. Previous study observed that when the annealing is insufficient, surface doping by potassium cannot induce well-defined superconductivity [32], which is in contrast to Fe deposition, implying that tuning the stoichiometry has more effects than charge doping. Magnetic exchange bias effect measurements provided evidence for antiferromagnetic order in as-grown FeSe films, which becomes absent after annealing [26]. The dissolve of magnetic order may be a hint to the spin fluctuation, one of the proposed pairing mechanisms for FeSe [29,33]. Furthermore, scanning tunneling spectroscopy has shown that Fe vacancies destroy the superconducting gap and induce bound states [34]. By removing these defects with annealing or Fe deposition, we minimize the scattering of the Cooper pairs. Multiple factors need to be considered in future studies to interpret the emergence of superconductivity in FeSe thin films.

Since its discovery in 2012, FeSe/STO has drawn great attention to its particular properties that are distinct from the bulk, *e.g.* the high $T_c$, the electron doping and the interface electron-phonon coupling. Our work reveals the similarity of FeSe thin film to the bulk in terms of stoichiometry-controlled phase transition. We have successfully tuned the composition of few-layer FeSe films on STO by selective deposition of Fe or Se. The as-grown or Se-rich phase has Fe vacancies up to 20% and is insulating, while the metallic/superconducting phase is close to stoichiometric, which can be achieved by either high temperature annealing or deposition of a proper amount of Fe. Only when proper stoichiometry is reached can interfacial effects produce the superconductivity with high $T_c$. Our work unveils the essence of the anneal and highlights the crucial role of stoichiometry in the properties of FeSe on STO.

This research was supported by Max Planck-UBC-UTokyo Centre for Quantum Materials, by Canada First Research Excellence Fund, Quantum Materials and Future Technologies Program, by Natural Sciences and Engineering Research Council of Canada (NSERC), and by Canada Foundation for Innovation (CFI). We thank Bruce A. Davidson for fruitful discussions.

[1] H. Hosono and K. Kuroki, Physica C Supercond. **514**, 399 (2015).
[2] F.-C. Hsu, J.-Y. Luo, K.-W. Yeh, T.-K. Chen, T.-W. Huang, P. M. Wu, Y.-C. Lee, Y.-L. Huang, Y.-Y. Chu, D.-C. Yan, and M.-K. Wu, Proc. Natl. Acad. Sci. USA **105**, 14262 (2008).
[3] Q.-Y. Wang, Z. Li, W.-H. Zhang, Z.-C. Zhang, J.-S. Zhang, W. Li, H. Ding, Y.-B. Ou, P. Deng, K. Chang, J. Wen, C.-L. Song, K. He, J.-F. Jia, S.-H. Ji, Y.-Y. Wang, L.-L. Wang, X. Chen, X.-C. Ma, and Q.-K. Xue, Chin. Phys. Lett. **29**, 037402 (2012).
[4] S. He, J. He, W. Zhang, L. Zhao, D. Liu, X. Liu, D. Mou, Y. B. Ou, Q. Y. Wang, Z. Li, L. Wang, Y. Peng, Y. Liu, C. Chen, L. Yu, G. Liu, X. Dong, J. Zhang, C. Chen, Z. Xu, X. Chen, X. Ma, Q. Xue, and X. J. Zhou, Nat. Mater. **12**, 605 (2013).
[5] S. Tan, Y. Zhang, M. Xia, Z. Ye, F. Chen, X. Xie, R. Peng, D. Xu, Q. Fan, H. Xu, J. Jiang, T. Zhang, X. Lai, T. Xiang, J. Hu, B. Xie, and D. Feng, Nat. Mater. **12**, 634 (2013).
[6] D. Huang and J. E. Hoffman, Annu. Rev. Condens. Matter Phys. **8**, 311 (2017).
[7] Z. Li, J. P. Peng, H. M. Zhang, W. H. Zhang, H. Ding, P. Deng, K. Chang, C. L. Song, S. H. Ji, L. Wang, K. He, X. Chen, Q. K. Xue, and X. C. Ma, J. Phys.: Condens. Matter **26**, 265002 (2014).
[8] C.-L. Song, Y.-L. Wang, Y.-P. Jiang, Z. Li, L. Wang, K. He, X. Chen, X.-C. Ma, and Q.-K. Xue, Phys. Rev. B **84**, 020503(R) (2011).
[9] W. Zhang, Z. Li, F. Li, H. Zhang, J. Peng, C. Tang, Q. Wang, K. He, X. Chen, L. Wang, X. Ma, and Q.-K. Xue, Phys. Rev. B **89**, 060506(R) (2014).
[10] Q. Wang, W. Zhang, Z. Zhang, Y. Sun, Y. Xing, Y. Wang, L. Wang, X. Ma, Q.-K. Xue, and J. Wang, 2D Mater. **2** (2015).
[11] W. Schuster, H. Mikler, and K. L. Komarek, Monatsh. Chem. **110**, 1153 (1979).
[12] T. M. McQueen, Q. Huang, V. Ksenofontov, C. Felser, Q. Xu, H. Zandbergen, Y. S. Hor, J. Allred, A. J. Williams, D. Qu, J. Checkelsky, N. P. Ong, and R. J. Cava, Phys. Rev. B **79**, 014522 (2009).
[13] A. J. Williams, T. M. McQueen, and R. J. Cava, Solid State Commun. **149**, 1507 (2009).
[14] C. Koz, M. Schmidt, H. Borrmann, U. Burkhardt, S. Rößler, W. Carrillo-Cabrera, W. Schnelle, U. Schwarz, and Y. Grin, Z. Anorg. Allg. Chem. **640**, 1600 (2014).
[15] S. Rößler, C. Koz, L. Jiao, U. K. Rößler, F. Steglich, U. Schwarz, and S. Wirth, Phys. Rev. B **92**, 060505(R) (2015).
[16] T. K. Chen, C. C. Chang, H. H. Chang, A. H. Fang, C. H. Wang, W. H. Chao, C. M. Tseng, Y. C. Lee, Y. R. Wu, M. H. Wen, H. Y. Tang, F. R. Chen, M. J. Wang, M. K. Wu, and D. Van Dyck, Proc. Natl. Acad. Sci. USA **111**, 63 (2014).
[17] F. Li, Q. Zhang, C. Tang, C. Liu, J. Shi, C. Nie, G. Zhou, Z. Li, W. Zhang, C.-L. Song, K. He, S. Ji, S. Zhang, L. Gu, L. Wang, X.-C. Ma, and Q.-K. Xue, 2D Mater. **3**, 024002 (2016).
[18] D. Huang, T. A. Webb, C.-L. Song, C.-Z. Chang, J. S. Moodera, E. Kaxiras, and J. E. Hoffman, Nano Lett. **16**, 4224 (2016).
[19] Y. Hu, Y. Xu, Q. Wang, L. Zhao, S. He, J. Huang, C. Li, G. Liu, and X. J. Zhou, Phys. Rev. B **97**, 224512 (2018).
[20] W. Zhao, M. Li, C.-Z. Chang, J. Jiang, L. Wu, C. Liu, J. S. Moodera, Y. Zhu, and M. H. W. Chan, Sci. Adv. **4**, eaao2682 (2018).
[21] W.-H. Zhang, Y. Sun, J.-S. Zhang, F.-S. Li, M.-H. Guo, Y.-F. Zhao, H.-M. Zhang, J.-P.

Peng, Y. Xing, H.-C. Wang, T. Fujita, A. Hirata, Z. Li, H. Ding, C.-J. Tang, M. Wang, Q.-Y. Wang, K. He, S.-H. Ji, X. Chen, J.-F. Wang, Z.-C. Xia, L. Li, Y.-Y. Wang, J. Wang, L.-L. Wang, M.-W. Chen, Q.-K. Xue, and X.-C. Ma, Chin. Phys. Lett. **31**, 017401 (2014).

[22] K. Zou, S. Mandal, S. D. Albright, R. Peng, Y. Pu, D. Kumah, C. Lau, G. H. Simon, O. E. Dagdeviren, X. He, I. Božović, U. D. Schwarz, E. I. Altman, D. Feng, F. J. Walker, S. Ismail-Beigi, and C. H. Ahn, Phys. Rev. B **93**, 180506(R) (2016).

[23] M.-L. Zhang, J.-F. Ge, M.-C. Duan, G. Yao, Z.-L. Liu, D.-D. Guan, Y.-Y. Li, D. Qian, C.-H. Liu, and J.-F. Jia, Acta. Phys. Sin. **65**, 127401 (2016).

[24] P. K. Maheshwari, R. Jha, B. Gahtori, and V. P. S. Awana, J. Supercond. Nov. Magn. **28**, 2893 (2015).

[25] Y. Fang, D. H. Xie, W. Zhang, F. Chen, W. Feng, B. P. Xie, D. L. Feng, X. C. Lai, and S. Y. Tan, Phys. Rev. B **93**, 184503 (2016).

[26] Y. Zhou, L. Miao, P. Wang, F. F. Zhu, W. X. Jiang, S. W. Jiang, Y. Zhang, B. Lei, X. H. Chen, H. F. Ding, H. Zheng, W. T. Zhang, J. F. Jia, D. Qian, and D. Wu, Phys. Rev. Lett. **120**, 097001 (2018).

[27] X. Liu, D. Liu, W. Zhang, J. He, L. Zhao, S. He, D. Mou, F. Li, C. Tang, Z. Li, L. Wang, Y. Peng, Y. Liu, C. Chen, L. Yu, G. Liu, X. Dong, J. Zhang, C. Chen, Z. Xu, X. Chen, X. Ma, Q. Xue, and X. J. Zhou, Nat. Commun. **5**, 5047 (2014).

[28] C. Liu, G. Wang, and J. Wang, J. Phys.: Condens. Matter **31**, 285002 (2019).

[29] D.-H. Lee, Annu. Rev. Condens. Matter Phys. **9**, 261 (2018).

[30] H. Zhang, D. Zhang, X. Lu, C. Liu, G. Zhou, X. Ma, L. Wang, P. Jiang, Q. K. Xue, and X. Bao, Nat. Commun. **8**, 214 (2017).

[31] T. Nishimura, A. Ikeda, H. Namba, T. Morishita, and Y. Kido, Surface Science **421**, 273 (1999).

[32] C. Tang, D. Zhang, Y. Zang, C. Liu, G. Zhou, Z. Li, C. Zheng, X. Hu, C. Song, S. Ji, K. He, X. Chen, L. Wang, X. Ma, and Q.-K. Xue, Phys. Rev. B **92**, 180507 (2015).

[33] Q. Wang, W. Zhang, W. Chen, Y. Xing, Y. Sun, Z. Wang, J.-W. Mei, Z. Wang, L. Wang, X.-C. Ma, F. Liu, Q.-K. Xue, and J. Wang, 2D Mater. **4**, 034004 (2017).

[34] C. Liu, J. Mao, H. Ding, R. Wu, C. Tang, F. Li, K. He, W. Li, C.-L. Song, X.-C. Ma, Z. Liu, L. Wang, and Q.-K. Xue, Phys. Rev. B **97**, 024502 (2018).

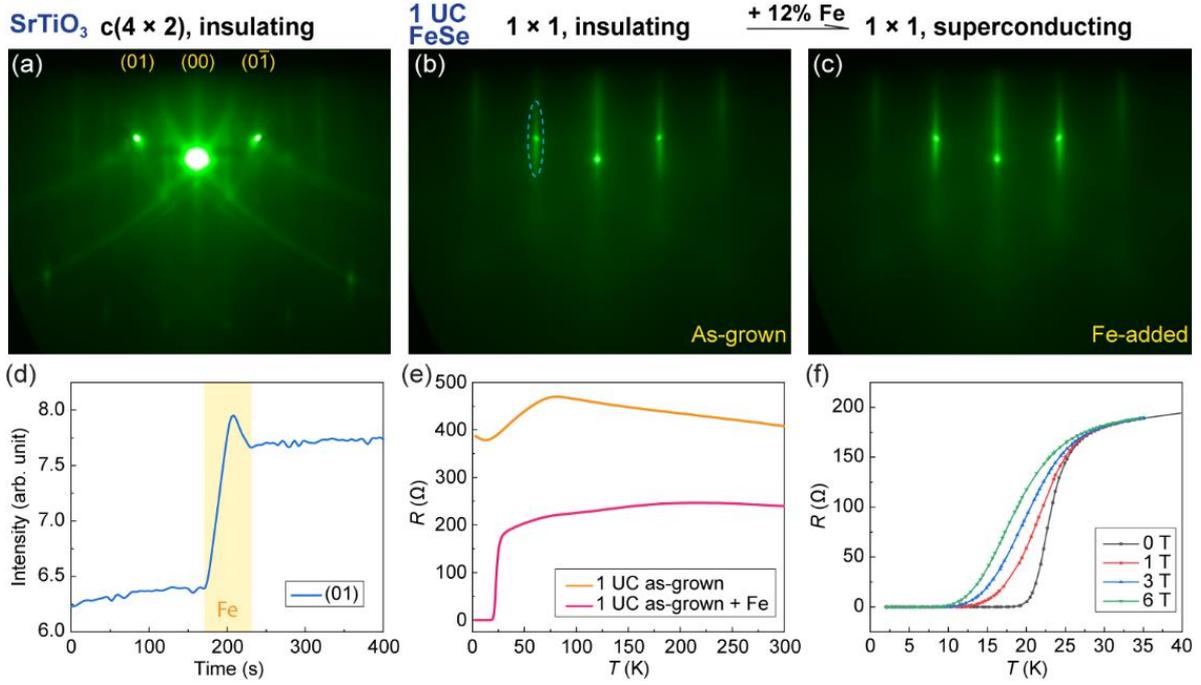

FIG. 1. (a) Reflection high-energy electron diffraction (RHEED) image of SrTiO$_3$ substrate with incident beam along [100] direction. (b) and (c) RHEED images of as-grown 1-UC FeSe film and the same sample after Fe deposition, respectively. (d) Integrated intensity of (01) diffraction streak as a function of time. The monitored areas are marked by the dashed ellipse in (b). The yellow interval indicates the period of Fe deposition. (e) Temperature dependence of the resistance of as-grown and Fe-adjusted monolayer FeSe films. (f) Temperature dependence of the resistance of Fe-adjusted monolayer FeSe film at low temperature with different out-of-plane magnetic field.

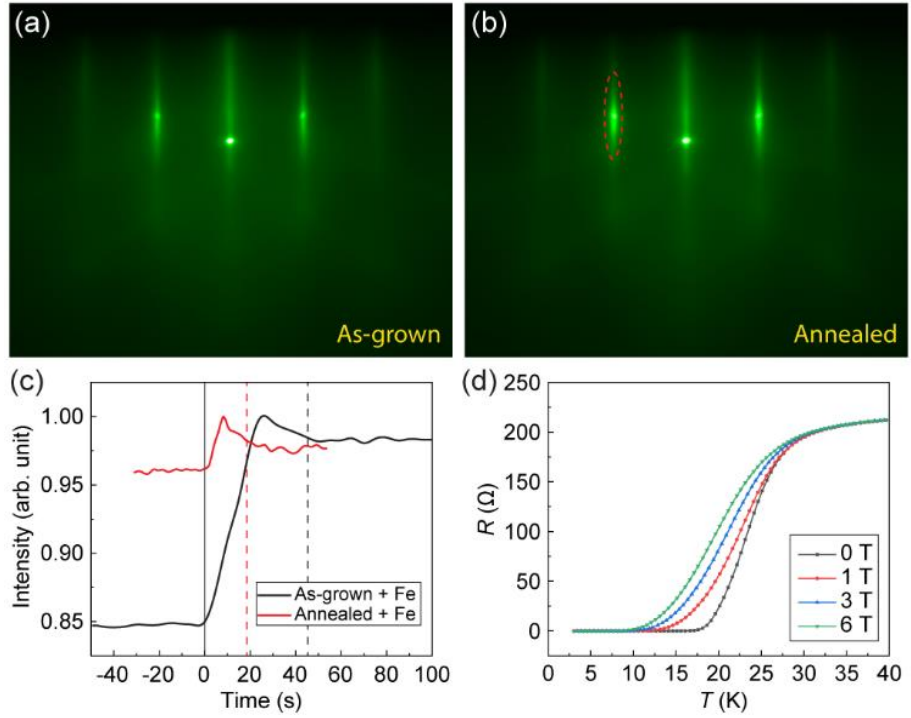

FIG. 2. (a) and (b) RHEED images of as-grown and annealed 1-UC FeSe film, respectively. (c) Integrated intensity of (01) diffraction streak of the annealed film in (b) (red) and an as-grown 1-UC film (black) as a function of time during Fe deposition. For each curve, the maximal intensity is rescaled to 1 and the starting time of Fe deposition is reset to 0. Each dashed line indicates the end time of Fe deposition for the curve with the same colour. (d) Temperature dependence of the resistance of the FeSe film in (b) after Fe deposition, with different out-of-plane magnetic field.

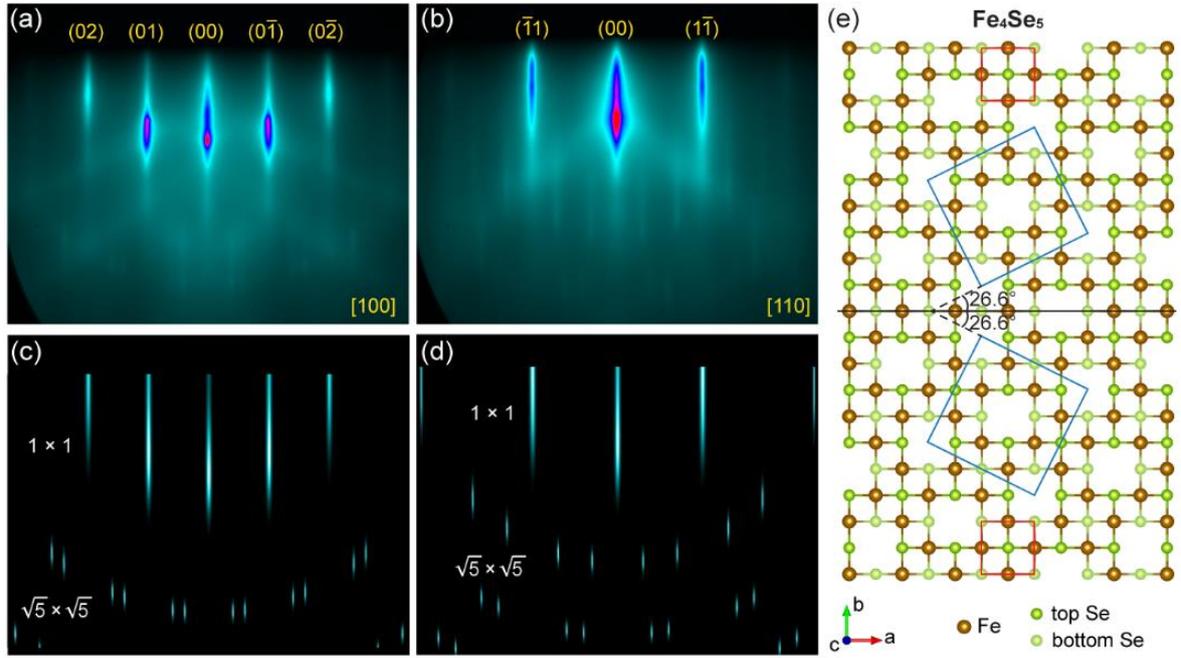

FIG. 3. (a) and (b) RHEED images of $\sqrt{5}\times\sqrt{5}$ 5-UC Fe$_x$Se film with the electron beam along [100] and [110] directions, respectively. (c) and (d) Simulated RHEED patterns of a $1\times 1$ square lattice with two $(\sqrt{5}\times\sqrt{5})$-$R26.6°$ supercells, with the electron beam in [100] and [110] directions, respectively. (e) Sketched lattice structure of Fe-vacancy phase Fe$_4$Se$_5$ with two possible domains [16]. The red and blue squares indicate the unit cells of FeSe and Fe$_4$Se$_5$, respectively.

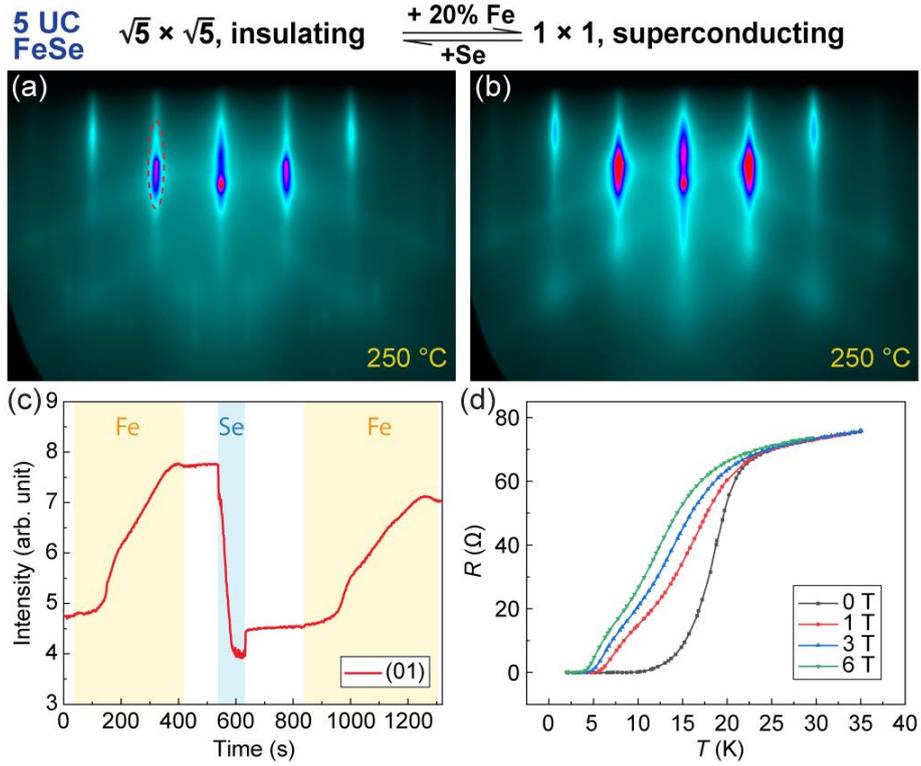

FIG. 4. (a) and (b) RHEED images of Se-added and Fe-adjusted 5-UC FeSe films, respectively. (c) Integrated intensity of (01) diffraction streaks in the elliptical area in (a), as a function of time. The yellow and blue intervals indicate the periods of Fe and Se deposition, respectively. The Se/Fe flux ratio was ~5. (d) Temperature dependence of the resistance of the FeSe film after two cycles of Se-Fe deposition, with different out-of-plane magnetic field.

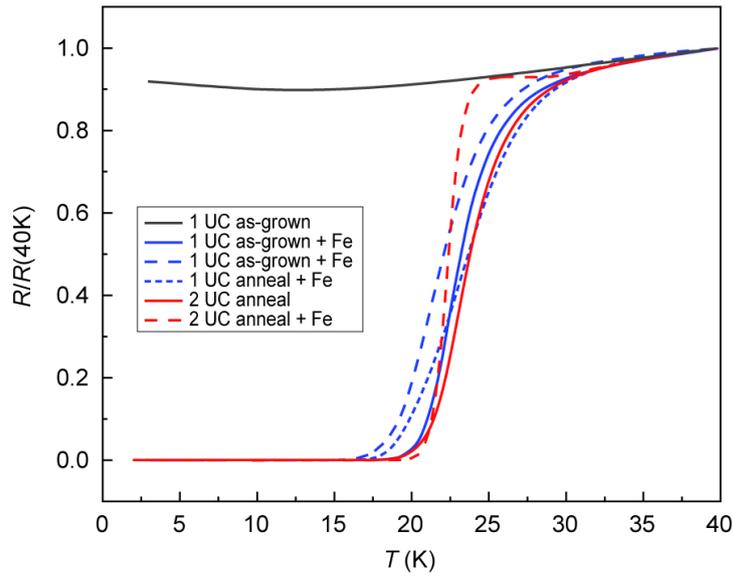

FIG. 5. Summary of transport measurement results of different samples in this work. The anneal was carried out at 480 °C for 3 hours. All samples were capped by 12-UC FeTe.